\documentclass[smallabstract,smallcaptions]{dccpaper}

\usepackage{subfig}
\usepackage{epsfig}
\usepackage{citesort}
\usepackage{amsmath}
\usepackage{amssymb}
\usepackage{comment}
\usepackage{diagbox}
\usepackage{color}
\usepackage{multirow}
\usepackage{url}

\usepackage{booktabs}
\usepackage{multirow}
\usepackage{subfig} 
\usepackage{rotating}
\usepackage{multicol}

\usepackage[utf8]{inputenc}
\usepackage{epstopdf}
\usepackage{tikz}

\newlength{\figurewidth}
\newlength{\smallfigurewidth}

\newcommand{\no}[1]{}

\begin{document}

\title
{{\large\textbf{Semantrix: A Compressed Semantic Matrix}}
	\footnote{$^{\S}$ {\em Funded in part by 
		European Union's Horizon 2020 research and innovation programme
		under the Marie Sklodowska-Curie grant agreement No 690941 (project BIRDS). 
		G.N. is funded 
		  by the Millennium Institute for Foundational Research on Data (IMFD), Chile. 
		The Spanish group is funded 
		  by Xunta de Galicia/FEDER-UE [CSI: ED431G/01 and GRC: ED431C 2017/58]; 
		  by Xunta de Galicia/GAIN [IN848D-2017-2350417];
		  by Xunta de Galicia Conecta-Peme 2018 [Gema: IN852A 2018/14];
		  by MCIU-AEI/ FEDER-UE [ETOME-RDFD3: TIN2015-69951-R, Datos 4.0: TIN2016-78011-C4-1-R,  BIZDEVOPS: RTI2018-098309-B-C32];
          and by FPI Program [BES-2017-081390] (T.V.R.).
   	   }
   }
}

\author{%
	Nieves R. Brisaboa$^{\ast}$, Antonio Fari{\~{n}}a$^{\ast}$, \\ Gonzalo Navarro$^{\dag}$, and Tirso V. Rodeiro$^{\ast}$\\[0.5em]
	{\small\begin{minipage}{\linewidth}\begin{center}
				\begin{tabular}{ccc}
					$^{\ast}$Universidade da Coru\~na & \hspace*{0.5in} & $^{\dag}$University of Chile \\
					Centro de Investigaci\'on CITIC, Database Lab && Department of Computer Science \\					
					A Coruña, Spain && IMFD \\
					\url{brisaboa@udc.es}, \url{fari@udc.es},&& Santiago, Chile \\
					\url{tirso.varela.rodeiro@udc.es}, && \url{gnavarro@dcc.uchile.cl}\\
				\end{tabular}
	\end{center}\end{minipage}}
}

\maketitle
\thispagestyle{empty}

\begin{abstract}
	We present a compact data structure to represent both the duration and length of homogeneous segments of trajectories from moving objects in a way that, as a data warehouse, it allows us to efficiently answer cumulative queries. The division of trajectories into relevant segments has been studied in the literature under the topic of Trajectory Segmentation. In this paper, we design a data structure to compactly represent them and the algorithms to answer the more relevant queries. We experimentally evaluate our proposal in the real context of an enterprise with mobile workers (truck drivers) where we aim at analyzing the time they spend in different activities. To test our proposal under higher stress conditions we generated a huge amount of synthetic realistic trajectories and evaluated our system with those data to have a good idea about its space needs and its efficiency when answering different types of queries. 
\end{abstract}

\section{Introduction}
 
In recent works, the need for analyzing trajectories in a higher abstraction level than the one offered by a sequence of GPS points, had led to the definition of the concept of {\em semantic trajectories} \cite{YanPSC10,Alvares:2007:MET:1341012.1341041,ParentSRAABDGMPTY13}. Basically, the idea is to split the trajectory of a mobile object into segments ({\em segmentation}) that are relevant according to some parameter (place, speed, activity, etc). 
After that, each segment is labeled with a semantically rich tag (``driving in slow traffic", ``shopping at the mall", ``working in customer facilities", ``working in the office", ``refueling", etc) \cite{Yan0PSA13,DodgeLW12,BrisaboaLuaces:w2gis2017,Mello2019MASTERAM}. 
Semantic trajectories should theoretically allow to analyze trajectories in a more relevant abstraction level. However, on the one hand, there are no standard ways to represent them, and on the other hand, the most relevant queries would need to accumulate the duration/length (size) of the semantically homogeneous segments. Therefore, there is an actual need for representing them in such a way that, as a data warehouse, it enables us to efficiently support queries.

The use of semantic trajectories has many applications. For example, it allows us to label the work/activity done by a mobile worker (e.g. a worker who moves/drives to visit customers) in each moment of the working day; 
or to know the state in which all the cars from a taxi company are (e.g. ``traffic jam'', ``normal traffic", ``stopped at semaphore", etc); or to classify a storm in different moments of its evolution.  In the context of our work, we deal with a set of trucks that collect organic waste from farms within a large area of Spain. 

Semantic trajectories are complex objects. They include spatio-temporal data representing the sequence of GPS points (polyline of the trajectory segment) and the actual time instants in which the mobile object went through the points of the trajectory. They also include a textual tag that identifies the place, activity speed, or whatever interesting aspect was used to separate a segment of the trajectory from the next segment. 

Nowadays, there is no standard (or at least an usual) way to represent those multidimensional complex data. Of course, we could use GIS\footnote{Geographic Information Systems} technology to define a trajectory segments table by  providing: a semantic trajectory ID, its geometry, the initial and final timestamps, the trajectory or object ID the segment belongs to, and the tag of the semantic trajectory. However, that GIS-based solution has two main problems: it uses too much space (as a consequence, the table would not fit into memory) and exploiting those data would become rather inefficient (not only they are in disk but also cumulative queries must be performed at query time, hence leading to a time-consuming solution).  Note that the historical set of semantic trajectories from all the workers during each day could become a big data problem. Therefore, defining an effective solution (in terms of space consumption and query time) to store and analyze those data is a relevant problem.
\medskip

In this work, 
we tackle a real problem of a truck fleet from a transport company where  there is interest in monitoring which activities are being done by each truck driver during a given time period, but also gathering activity patterns is of interest. We present {\em semantrix}, a compact representation for the sequence of tags associated to semantic trajectories (that in our experimental data would represent activities) in such a way that we could efficiently answer different relevant types of queries oriented to analyze the data, particularly focusing on aggregated queries. Our proposal uses compact data structures based on bitmaps and sums-matrices to store the data in a pre-computed way (as it is usual in data warehouses). 

Note that we do not tackle the problem of labeling the semantic trajectories. In different contexts, such labeling process can be done either manually by the worker himself or automatically using detection strategies or machine learning. In our case, we followed the automatic approach in \cite{BrisaboaLuaces:w2gis2017}. In any case, we assume that trajectories have been previously split into homogeneous pieces and each segment has been correctly labeled.  In the same way, we do not deal with the spatial representation of trajectories (or segments of trajectories) as we assume that both the representation of the geometries and (in general) the representation of the cartography is done with a GIS. Therefore, our problem focuses in how to deal with the labels of the segments (the semantic information of the trajectories) and how long this segment lasts (its size/duration). Note that by knowing that a semantic trajectory $f_i$ from a mobile object $O_j$ lasts from an initial to an ending timestamp we can easily map that label over the corresponding geographic representation of the segment. Recall each segment in the spatial database has its initial and ending timestamp.

In consequence, the target of our representation is to enable the efficient exploitation of the semantic labels when dealing with queries such as: ``how many hours did my workers spend at refueling during the last month?", ``how many miles in average did my workers drive to meet customers?", or “how many of my workers had lunch between 14 and 15pm?”. In addition, we can also solve queries about the sequence/patterns of activities performed by a truck-driver: e.g. ``How many times (or who/when did) the activity {\em driving out of the planned route} was performed just after the activity {\em driving in slow traffic on the planned route}?”.

\section{Basic concepts} \label{basicConcepts}

In this section, we briefly describe some well-known data structures that make up the basic components of our proposal.

\begin{itemize}

\item \textbf{Bitvectors.} Bitvectors are the basic components of many Compact Data Structures. A bitvector $B[1,n]$ is a sequence of zeroes and ones of lenght $n$. The following operations are expected to be supported:

\begin{itemize}    
    \item $rank_1(B,i)$ returns the number of set bits in $B[1..i]$. Alternatively, $rank_0(B,i) = i - rank_1(B,i)$ and also $B[i] = rank_1(B,i) - rank_1(B,i-1)$.
    \item $select_1(B,i)$ returns the position in $1..n$ where the $i$th 1 occurs. Therefore, $rank_1(B,select_1(B,i)) = i$.
\end{itemize}

These operations can be supported in constant time by using $o(n)$ extra bits \cite{Jac89,Mun96}. There also exist compressed bitvector representations of $B$ \cite{Raman:2002:SID:545381.545411,okanohara2007practical,Golynski2007} that still support those operations and also permit to solve $access(B,i)$, which returns the original value $B[i]$.

\item \textbf{FM-index.} Given a text $T[1,n]$ built on an alphabet $\Sigma = [1,\sigma]$, the FM-index \cite{FerraginaM00} provides a self-indexed representation of $T$ based on the BWT \cite{BWT1994} of $T$ and the use backward searching for identifying pattern occurrences. It requires $5n H_k(T)+o(n)$ bits of space and permits to search for the occurrences of a pattern $P[i,m]$ in time $O(m+ occ~ log ^{1+\epsilon}n)$ ($occ$ being the number of occurrences of $P$ within $T$). Several variants of this scheme exist \cite{Ferragina01,Ferragina05,Ferragina07,Makinen05} which induce different  time/space tradeoffs for the counting, locating, and extracting operations.

\item \textbf{Summed Area Tables.} The Summed Area Tables were first introduced in computer graphics \cite{crow1984summed} to speed up  mipmapping. Given a matrix $A[1,r][1,c]$, for which we want to solve the operation $countRange(A,[x_1,y_1],[x_2,y_2])$ $ \leftarrow \sum_{i=x_1}^{x_2} \sum_{j=y_1}^{y_2} A[i][j]$, the  key idea of this approach is to create a new matrix $M[0,r][0,c]$ where all the cells in both row $0$ and column $0$ are set to zero, and any other cell $M[x][y]$ stores the total sum of all the previous cells within $A$ (to the left and up); i.e. $M[x][y] \leftarrow  \sum_{i=1}^{x} \sum_{j=1}^{y} A[i][j]$. An example showing matrices $A$ and $M$ is depicted in Figures~\ref{fig:accumM}(a) and ~\ref{fig:accumM}(b). Using $M$ allows us to solve $countRange$ operation in $O(1)$ time as $countRange(A,[x_1,y_1],[x_2,y_2]) \leftarrow {M[x_2,y_2]} - {M[x_2,y_1-1]} - {M[x_1-1,y_2]} + {M[x_1-1,y_1-1]} $. Basically, from a geometric point of view,  Figure~\ref{fig:accumM}(c) shows that to compute $countRange(A,[3,2],[7,4])$ we subtract from $M[7,4]=64$ (sum of all the values in $A[1,7][1,4]$) both the values in the area depicted with horizontal bars ($M[7,1] = 19 =$ sum of values in $A[1,7][1,1]$) and those  values in the area depicted with vertical bars ($M[2,4] = 18 =$ sum of values in $A[1,2][1,4]$). Since we are subtracting  the sum of values in the area depicted with both vertical and horizontal lines twice ($M[2,1] = 6 =$ sum of values in $A[1,1][2,1]$) we still have to add that value ($M[2,1] = 6$) once. Consequently, we obtain $countRange(A,[3,2],[7,4]) \leftarrow 64-19-18+6 = 33$.

\begin{figure}[ht!]
\begin{center}
  \includegraphics[scale=0.65]{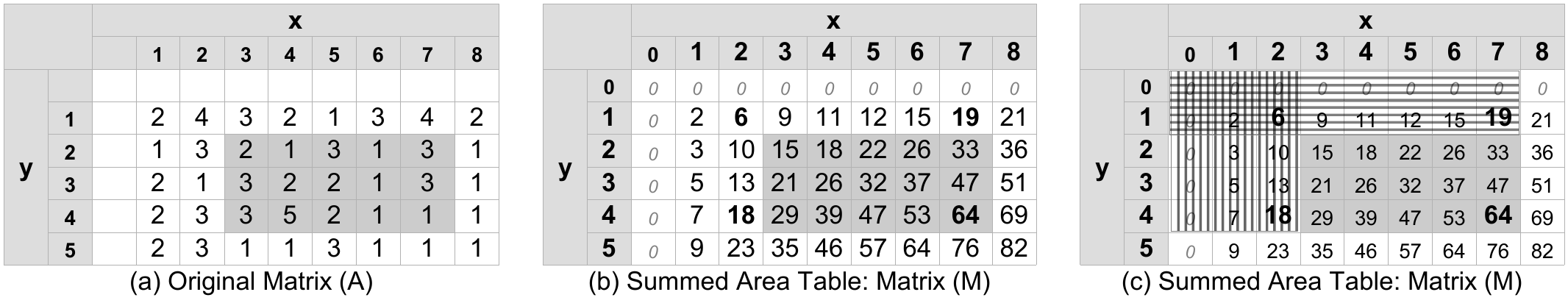}
  \caption{Summed Area Tables example.}
  \label{fig:accumM}
\end{center}
\end{figure} 
\vspace{-0.5cm}
\end{itemize}

\section{Our proposal: Semantrix }\label{sec:proposal}

\begin{figure}[!hbp]
  \centering
  \begin{minipage}[b]{0.3\textwidth}
  \begin{center}
      
  \includegraphics[scale=0.45]{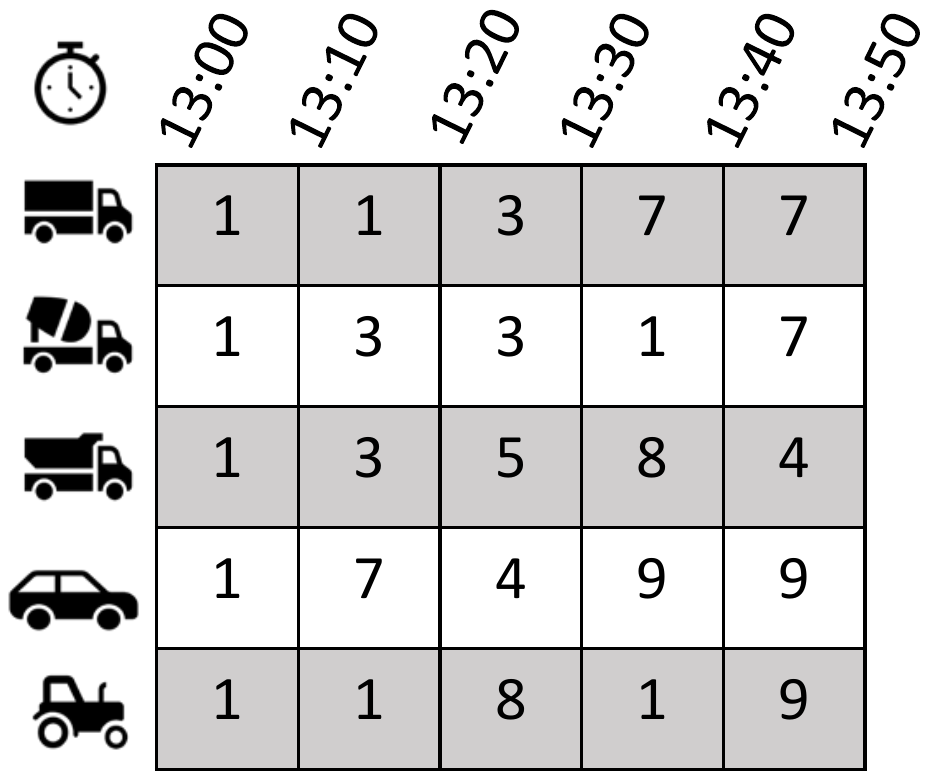}
  \caption{Naive matrix.}
  \label{fig:naivematrix}
  \end{center}
  \end{minipage}
  \begin{minipage}[b]{0.65\textwidth}
    \includegraphics[scale=0.45]{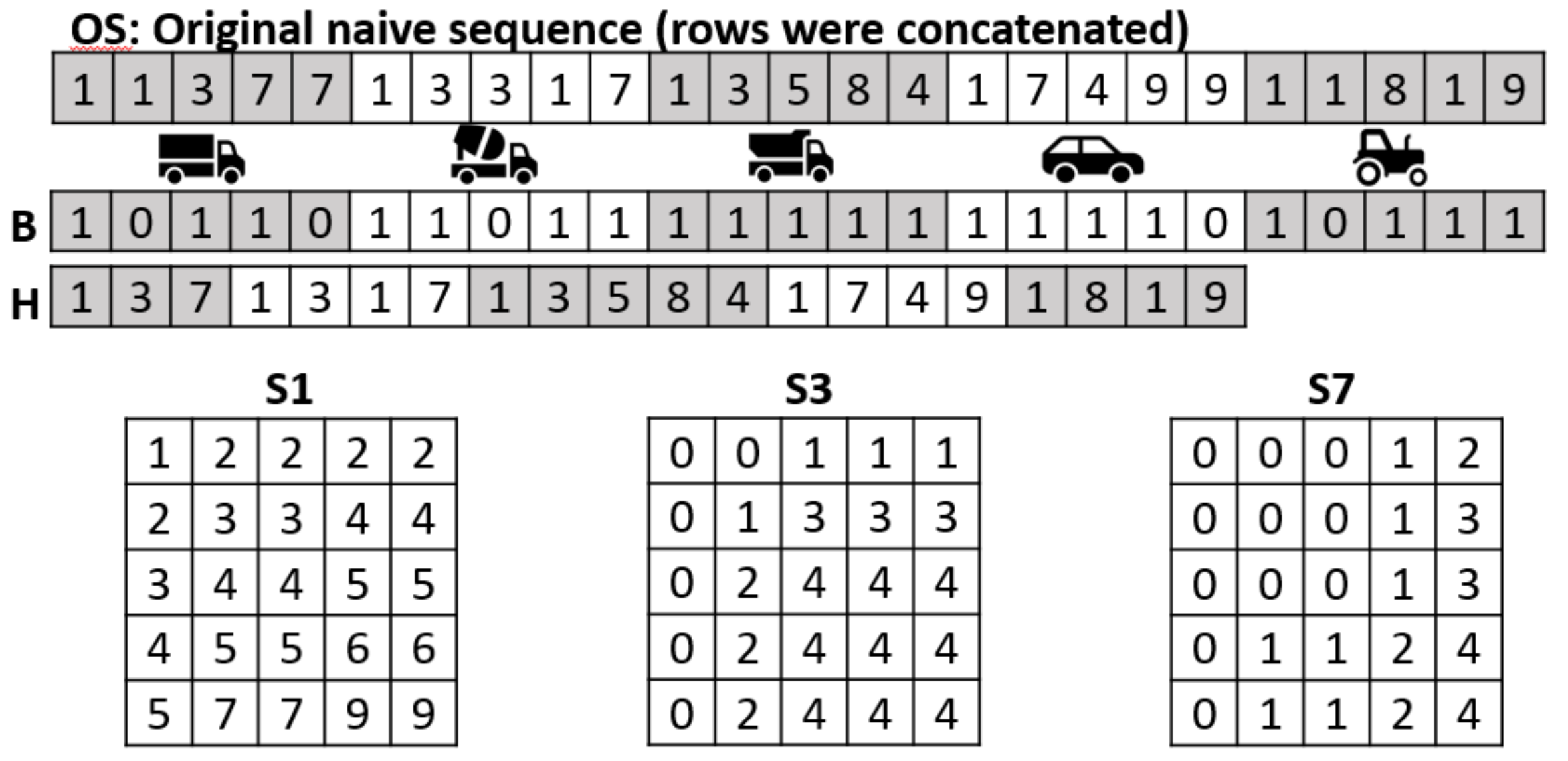}
    \caption{Semantrix structure.}
    \label{fig:semantrix}
  \end{minipage}
  
\end{figure}

In this work, we aim at creating a compressed representation of a set of semantic trajectories/activities in such a way that we could still answer different relevant queries efficiently. Particularly, we are targeting at aggregated queries. Note that our set of semantic trajectories can be gathered from the movements of several objects/vehicles along time.
A rather straightforward ({\em naive}) approach would be a solution based on a classic matrix where columns represent a discretization of the time in such a way that each column corresponds to a time interval related to the actual continuous time between two discretized time instants (e.g. 13:00 - 13:10). The rows represent each of the moving objects of study. Thus, a cell within this matrix contains the identifier of the (most-representative) activity performed by a given mobile object at a  particular time interval. For example, in the matrix in Figure~\ref{fig:naivematrix}, the car was performing the activity with id 4 from 13:20h to 13:30h.

\bigskip

\paragraph{Semantix structure:}

With the aim of improving the previous solution, we have created a new structure named \textit{semantrix} that represents all the information included in the previous {\em naive} original matrix, and considerably reduces pattern matching and aggregated queries times. This new structure encompasses three vectors: a bitvector $B$, an integer vector $H$, and a vector of matrices $S$. The former two structures permit us to compactly represent the original sequence of activities within the {\em naive} matrix. The later vector keeps one activity matrix for each possible activity so that, for each activity, it handles aggregated information for each vehicle and time-interval. Those structures, that are discussed below, are depicted in Figure~\ref{fig:semantrix}.

\begin{itemize}

\item {Representing the {\em naive} matrix: \textbf{Bitvector $B$ and vector $H$}.} Recall the information that regards the activity performed by each mobile object during each discretized time interval was stored in the {\em naive} matrix previously. In addition, a given row $i$ ($1\leq i \leq r$) keeps particularly the activities for the $i$-th mobile object during each of the $I$ time intervals. Those $r$ rows can be concatenated to make up a unique sequence of rows ($OS[1,rI]$) as depicted in the top of Figure~\ref{fig:semantrix}. Note that since all those $r$ rows have the same length $I$ (number of time intervals) we retain the same direct-access capabilities as in the {\em naive} matrix. Yet, we also have the same space needs. To compactly represent $OS$ we use: 

 {\em i)} A {\bf bitvector $B[1,rI]$} aligned with $OS$ where we set a $1$ each time an activity switch occurs in $OS$; i.e. we set $B[1]=0$ and then, $\forall i \in\{2..rI\}$ we set $B[i]= 1$ if $OS[i] \neq OS[i-1]$;  we set $B[i]= 0$ otherwise. Finally, we also set a $1$ at positions $B[1+kI]~ \forall k=\{1..r\}$ to mark a row/mobile-object switch.

 {\em ii)} An {\bf integer vector $H[1,o]$}, such that $o= rank_1(B,rI)$ is aligned with the $o$ ones in $B$, and stores the $ids$ of the activities from $OS$ associated to those ones in $B$. Therefore, $\forall i \in \{1,o\}$ we set $H[i] = OS[select_1(B,i)]$. Note that $H$ contains, for each mobile object, a sequence with the identifiers of the activities it performed.

\item Storing aggregated information related to each activity: \textbf{Vector of Activity matrices $S$}. We have included a vector of matrices (one per activity) that operates as a kind of a classic data warehouse. The goal is to have cumulative information pre-computed to efficiently solve aggregated queries. Thereby, this vector $S$ contains one matrix ($S_i$) for each possible activity in the system having the data in each matrix pre-computed as in a \textit{Summed Area Table} (see section \ref{basicConcepts}). In Figure \ref{fig:semantrix}, it is shown how the cumulative activity matrices $S_1$, $S_3$, and $S_7$ for the activities $1$, $3$, and $7$ in our working example would look like (note that we are not showing the content of the other $S_i$ matrices). By using the $countRange$ operation we will be able to gather, for example, the number of times an activity was performed during a given time window.

\end{itemize}

\section{Supporting activities-related queries  in \textit{semantrix}}

In our scope, we can distinguish among three main types of queries. We found {\em individual queries} that aim at gathering the content of one particular cell from the original {\em naive} matrix (e.g. \textit{Which was the activity performed by a given  mobile object $O_j$ at a given time instant $I_i$?}, or \textit{Which is the list of activities performed by a given mobile object $O_j$ during a given time interval $[I_s..I_e]$?}). There are also queries focusing on detecting if a given pattern of activities occurred (e.g. \textit{How many times the activity $A_i$ was followed by the activity $A_j$?}). Finally, we also have to deal with aggregated queries aimed to unravel the total values hidden within the matrix (e.g. \textit{How much time was actually spent by all the mobile objects while performing the activity $A_i$ during a given time window $[I_s, I_e]$}). To support this types of operations we used the different structures within \textit{semantrix}.

\begin{itemize}
    \item \textbf{Individual queries:} These kind of queries are easily solved just using the bitvector $B$ and vector $H$. First, with a $rank$ operation over the bitmap we obtain the position(s) of interest; and then this position is used to access  $H$ to retrieve the activity/ies within the particular time window.

    \item \textbf{Pattern queries:} For these queries, a FM-index built on top of $H$ vector is used. Therefore, we use its self-indexing capabilities to efficiently locate patterns of activities. Particularly, to solve query ``How many times was activity $A_i$ followed by $A_j$?" we simply rely on $count(A_iA_j)$ over the FM-index of $H$.

    \item \textbf{Aggregated queries:} With the help of the activity matrices ($S_i$) and the $countRange$ operation, most aggregated queries can be solved in constant time.

\end{itemize}

\section{Experimental evaluation}

It is worth recalling that the seminal idea for this work arose as a recent project shared with a local company devoted to the transportation of organic waste. Accordingly, the actual experimental evaluation is now taking place on a real environment. Our system is being used on a daily basis to manage the activities of the trucks from the enterprise. The relevant activities for this company we have to deal with are:

\begin{multicols}{2}
\begin{itemize}\setlength\itemsep{-1px}
    \item $A_1$ Being at headquarters
    \item $A_2$ Working at a customer place
    \item $A_3$ Normal transit on planned route
    \item $A_4$ Slow transit on planned route
    \item $A_5$ Normal transit out of planned route
    \item $A_6$ Slow transit out of planned route
    \item $A_7$ Taking a break
    \item $A_8$ Undefined/unknown activity
    \item $A_9$ Inactive
    
\end{itemize}
\end{multicols}

We present experiments comparing our proposal {\em semantrix} with other representations and show both the space needs and their performance at query time. Below, we discuss the baseline representations used, we present our test dataset and finally we show the corresponding experimental results.

\subsection{Representations compared with {\em semantrix}: {\em naive}, {\em baseline+}, and {\em Diff}}
We have included in our experiments the {\em naive} original matrix discussed in Section~\ref{sec:proposal}. 

\begin{figure}[ht!]
\centering
  \includegraphics[width=0.75\textwidth]{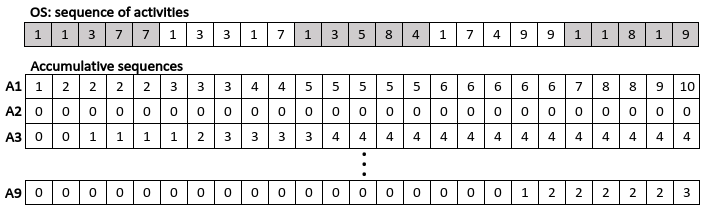}
  \caption{Baseline+ example}
  \label{fig:baseline}
\end{figure}

Additionally, we have implemented a more elaborated baseline named {\em baseline+} (see Figure \ref{fig:baseline}). It is based on the sequence of all the activities performed ordered both chronologically, and by moving object. It consists basically on the $OS$ vector (i.e. sequence composed of the rows from the original matrix). Yet, we have also included a set of aggregated sequences to boost solving aggregated operations. There is one sequence per activity that gathers all the cumulative data in chronological order.
Thereby, \textit{individual} and \textit{pattern} queries are solved with the activity sequence, while the cumulative sequences deal with the data warehouse-like queries. 

\bigskip 

Finally, we also implemented a variant of \textit{semantrix} following the ideas presented in \cite{BrisaboaFGRR18} named (\textit{Diff}) where the activity matrices $S_i$ are represented in a slightly more compact way. The idea is to sample some rows and to represent the non-sampled rows as differences with respect to the closest sample and the actual value. This implies a space/time trade-off. 
\subsection{Datasets}

Since our system has not been used over a relevant amount of time (6 months or more) yet, there are not enough real data to test our proposal in a real environment. Nonetheless, we have generated a synthetic dataset according to the actual constraints and the current existing statistics, where we have recreated realistic information about daily truck activities in the company. We have discretized the time using 5-min intervals, which is a sensible time lapse considering the speed of the trucks. We assume a small company that has $20$ trucks that work $8$ hours every day of the week. Assuming those preconditions and the nine activities discussed above, three datasets with different temporal sizes were created: one month ($12 * 8 * 7 * 4 = 2688$ time instants), six months ($12 * 8 * 7 * 4 * 6 = 16128$ instants) and one year ($12 * 8 * 7 * 4 * 12 = 32256$ instants).

\subsection{Experiments: space and query time comparison}
\begin{figure}[ht!]
\begin{center}
  \includegraphics[width=0.45\textwidth]{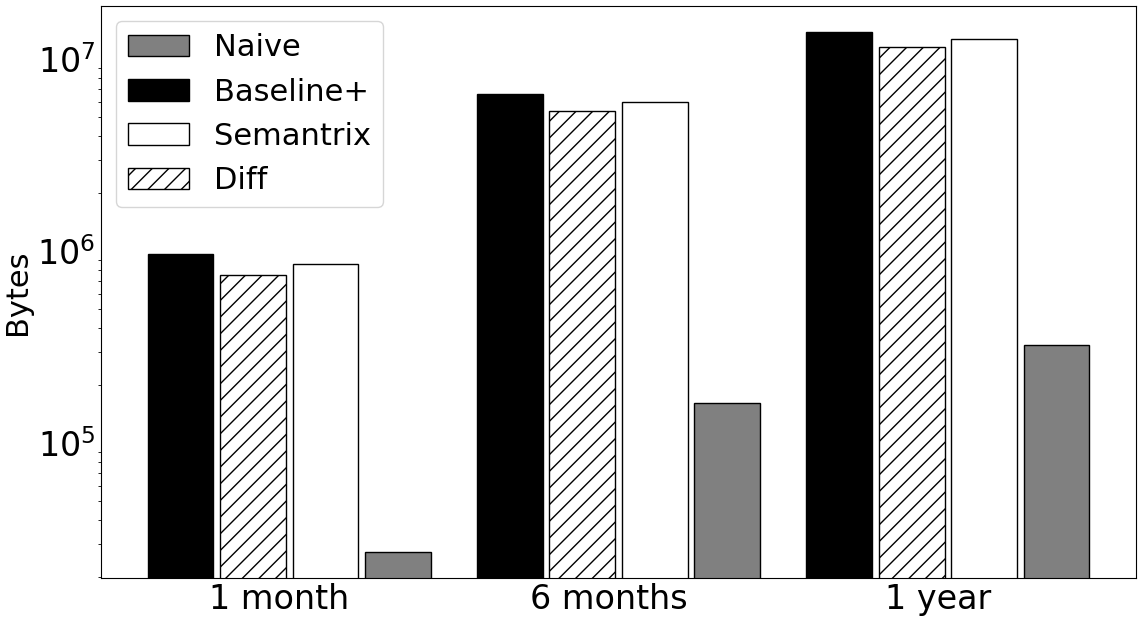}
  \caption{Space measurements}
  \label{fig:space}
\end{center}
\end{figure}

We have compared the space requirements of the tested techniques.  As shown in Figure \ref{fig:space}, the original matrix ({\em naive)} needs, by far, less space as it only stores the activity values within the original matrix. 

The others use roughly the same space. Yet, it is worth noting that {\em Diff} (sampling every 4 rows) requires around $15$\% less space than {\em semantrix}. {\em Baseline+} uses around $8$\% less space than {\em semantrix}.

To test query performance, we have chosen one query of each type (we have skipped the results from single-query type due to space constraints. Yet, the results showed only negligible differences among all the techniques). For pattern-queries we used the query {\em ``How many times was activity $x$ followed by activity $y$?"}, and for aggregated-queries, we used the query {\em ``how many times were trucks 1,2, and 3 performing activity $x$ from 11am to 12pm ($12$ time Intervals)?"}. We have measured average execution times from $10,000$ randomly generated queries on an Intel(R) Core(TM) i7-3820 CPU @ 3.60GHz machine running Debian GNU/Linux 9.9. Our implementations use components from the SDSL Library\footnote{https://github.com/simongog/sdsl-lite}. The compiler used was g++ 6.3.0.

\begin{figure}[ht!]
\begin{center}
  \includegraphics[width=0.45\textwidth]{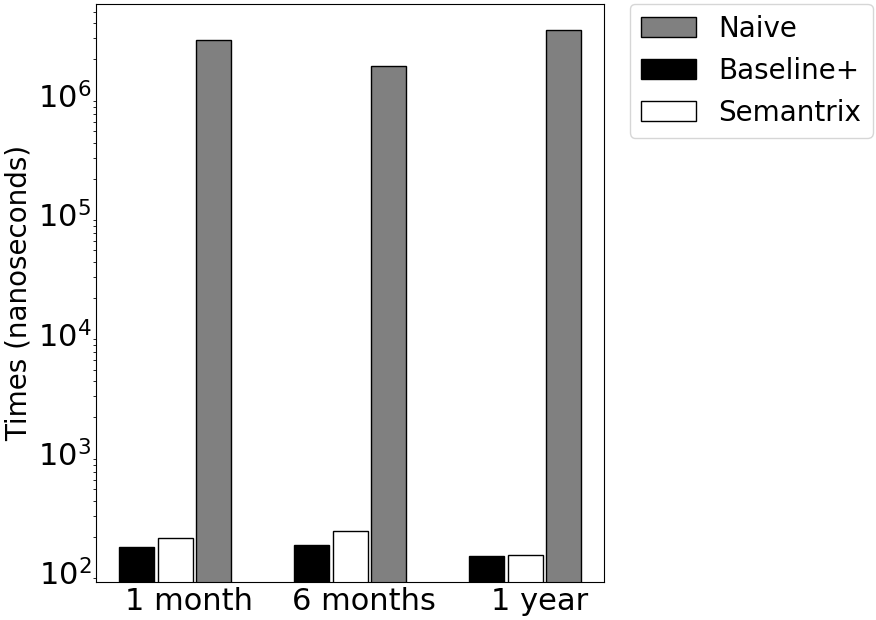}
  \includegraphics[width=0.45\textwidth]{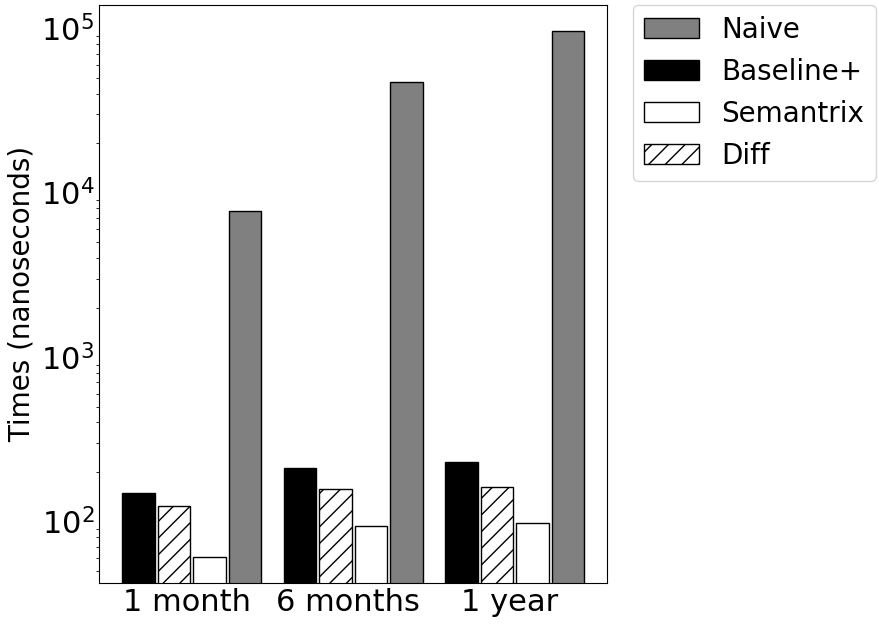}
 
  \caption{Times for pattern queries (left), and times for aggregation queries (right)}
  \label{fig:pattern_graph}
\vspace{-5mm}      
\end{center}
\end{figure}

We can see in Figure~\ref{fig:pattern_graph} that the techniques recreating a data warehouse ({\em semantrix}, {\em Diff}, and \textit{baseline+}) are much faster solving both pattern and aggregation queries than {\em naive}. Actually, {\em naive} approach, which must traverse the original matrix, becomes several orders of magnitude slower than both \textit{semantrix}, and \textit{baseline+} when solving pattern queries (Figure~\ref{fig:pattern_graph}.(left)). However, these latter structures obtain similar results as they both rely on an FM-index to solve pattern queries (with the difference that \textit{semantrix} needs an additional access to $B$).

For aggregated queries, Figure~\ref{fig:pattern_graph}.(right) shows, as expected, that {\em semantrix} is clearly the fastest technique. As in the previous experiment, {\em naive}  needs to explore the whole queried submatrix, whereas  \textit{semantrix} counterparts and \textit{baseline+} benefit from their aggregated data. In this case, {\em Diff} is around $40$\% slower than \textit{semantrix}.

\section{Conclusions and future work}

We have analyzed the problem of representing and managing semantic trajectories  in a compact and efficient way.  We present a data structure named {\em semantrix} to handle a semantic data warehouse for the trajectories from mobile objects, and we show how it supports  different types of queries. The proposal works on top of the compressed activity sequence (ordered chronologically and by mobile-object identifier) which leans on a bitmap for individual and pattern-matching queries. To improve the resolution of aggregated queries a cumulative matrix for each activity was appended to our structure enabling it to solve most accumulated queries in constant time. We have experimentally evaluated the proposed solution using realistic synthetic data that represent the truck movements of a real company. As a quality proof, it is worth recalling that our system is being used as part of a real company project, solving a real life problem.

\bigskip

Regarding future work, the first step will be to increase the scope of this work in order to represent in a compact way also the geometry of each semantically tagged segment or semantic trajectory. This idea opens a wide new field of possibilities to  perform queries combining spatial, temporal, and semantic constraints.

\Section{References} 
\bibliographystyle{IEEEbib}
\bibliography{refs}

\end{document}